\documentclass[aps,prl,twocolumn,showpacs,amsmath,amssymb]{revtex4-1}

\usepackage{graphicx}
\usepackage[bookmarks,bookmarksopen,bookmarksnumbered,colorlinks,linkcolor=red,linktocpage,citecolor=blue,urlcolor=cyan,pdfpagemode=UseOutline]{hyperref}
\usepackage{epstopdf}

\def\bea{\begin{eqnarray}}
\def\eea{\end{eqnarray}}
\def\ben{\begin{equation}}
\def\een{\end{equation}}
\def\benu{\begin{enumerate}}
\def\enu{\end{enumerate}}

\def\bei{\begin{itemize}}
\def\eei{\end{itemize}}
\def\benu{\begin{enumerate}}
\def\enu{\end{enumerate}}

\def\n{n}

\def\sss{\scriptscriptstyle\rm}

\def\1var{(\bx_1...\bx\N)}

\def\br{{\bf r}}

\def\bx{{x}}

\def\N{_{\sss N}}

\def\sph_int{ {\int d^3 r}}

\newcommand{\intd}{\mathrm{d}}
\newcommand{\vect}[1]{\mathbf{#1}}

\providecommand{\abs}[1]{\left|#1\right|}

\DeclareMathOperator{\erfc}{erfc}

\def\calE{{\cal E}}
\def\brp{\bar\br}

\begin{document}

\title{The effect of cusps in time-dependent quantum mechanics}
\author{Zeng-hui Yang}
\affiliation{Department of Physics and Astronomy, University of Missouri - Columbia, MO 65211, USA}
\author{Neepa T. Maitra}
\affiliation{Department of Physics and Astronomy, Hunter College of the City University of New York, NY 10065, USA}
\author{Kieron Burke}
\affiliation{Department of Chemistry, University of California - Irvine, CA 92697, USA}
\date{\today}

\begin{abstract}
Spatial cusps in initial wavefunctions can lead to non-analytic behavior in time.
We suggest a method for calculating the short-time behavior
in such situations.  For these cases, the density does not match its Taylor-expansion in time, but
the Runge-Gross proof of time-dependent density functional theory still holds, as it requires only the potential to be time-analytic. 
\end{abstract}

\maketitle

Within the Born-Oppenheimer(BO) approximation, a vast amount of useful
information (such as geometry, thermodynamics, and vibrations) of a molecule or
solid can be extracted from the electronic ground state.
But electronic excitations are important for many areas,
from photochemistry to photoemission spectra\cite{EFB09}.  Ground-state density
functional theory(DFT)\cite{HK64,KS65,FNM03} has been very successful
for the former problem, and its time-dependent analog (TDDFT)\cite{RG84}
has become popular for the latter\cite{MUNR06}.  Within linear response, 
TDDFT yields useful predictions for the excited states of many molecules\cite{EFB09},
and extensions to solids are a keen area of research\cite{ORR02,BSSR07}.

All density functional theories rely on a one-to-one 
correspondence between the one-body potential, such as $-Z/r$ for an atom,
and the density $\n(\br)$, under some restrictions.  For ground-state DFT,
the particle statistics and interaction are fixed.
For the time-dependent case, one
also specifies the initial wavefunction.  The original proof in
the ground-state case of Hohenberg and Kohn\cite{HK64} has been
refined over the decades\cite{L79,L82,L83},
but its essence remains unchanged. 
In the time-dependent problem, after pioneering works by others\cite{HD82,H77,MD82,Y74,ML75,DG82,B84}, Runge and Gross\cite{RG84}
gave a proof assuming the one-body potential is time-analytic, i.e., equals its Taylor expansion in time around the initial time, for a finite time-interval. 
Despite recent attempts\cite{NS87,L01,MTWB10,RL10,T10}, no generally applicable proof has
been found that avoids this expansion.

Modern DFT calculations employ the Kohn-Sham scheme\cite{KS65}, in which a fictitious set of
non-interacting electrons reproduces the one-electron density of the 
real system.   If we assume that time-analytic potentials yield time-analytic
densities, van Leeuwen\cite{L99} gave a constructive procedure
for finding the TD KS potential for a given system,
a problem that has not yet been generally solved for the
ground-state case.
But, in the usual treatment, matter has cusps in its ground-state electronic wavefunctions
at the nuclei, and the resulting spatial non-analyticities are coupled
to the time-dependence in the TD Schr\"odinger equation\cite{MTWB10}.  Even in the most mundane example,
a hydrogen atom in a suddenly switched
electric field, this coupling leads to non-analytic short time dynamics, and its density is not time-analytic.

We develop a method for extracting the exact non-analytic short-time behavior of the Schr\"odinger equation
in the presence of cusps.  
There are distinct spatial regions with different asymptotic behavior for short times,
and by `asymptotic behavior' we mean a series expansion around the initial time $t=0$,
whose error vanishes as $t\to0_+$.
We calculate the exact short-time behavior for a hydrogen atom in an electric field, 
and demonstrate agreement with linear response theory in the limit of weak fields.
The constructive procedure 
for the
TD KS potential fails for our examples.
(In DFT-speak, the $v$-representability question is not solved.)
Nonetheless, the original proof of Runge and Gross remains valid.

\def\TE{^{\rm TE}}
Consider a single particle prepared in a field-free ground-state, $\psi_0$,
placed in a static field which is turned on at t=0, and remains on indefinitely.
The  exact TD wavefunction is
\ben
\psi(\br,t)=\sum_j c_j\exp(-i\epsilon_j t)\phi_j(\br),
\label{psi}
\een
where $\epsilon_j$ and $\phi_j(\br)$ are the
eigenvalues and functions of the Hamitonian operator $\hat H$, and $c_j=\langle \phi_j | \psi_0 \rangle$.
We define the time-Taylor series
\ben
\psi\TE(\br,t)=\sum_{p=0}^\infty c_p(\br)\, t^p,\quad
c_p(\br)=\sum_n c_n \frac{(-i\epsilon_n)^p}{p!} \phi_n(\vect{r}),
\label{psiTE}
\een
which is the 
result of the usual practice of Taylor-expanding (TE) the time-evolution operator:
\ben
\hat{U}(t)=1-i\hat{H}t-\hat{H}^2t^2/2+\cdots.
\label{eqn:UTE}
\een
In many cases,
the solutions agree,
but not necessarily
when the wavefunction has non-analyticities in space.
Although $\psi\TE$ \textit{formally} solves the TDSE, it may not be a valid solution.
Holstein and Swift\cite{HS72} reported failure of the TE solution in a
model system, in which $\psi_0$ has compact support
(all space-derivatives of $\psi_0$ vanish at the boundary).
However, such cases do not occur in routine electronic structure calculations, and 
may not be of concern in practice.  For the rest of this paper, we focus on 
the ubiquitious cusp in $\psi_0(\br)$ at a nucleus.

\begin{figure}[htbp]
\includegraphics[width=\columnwidth]{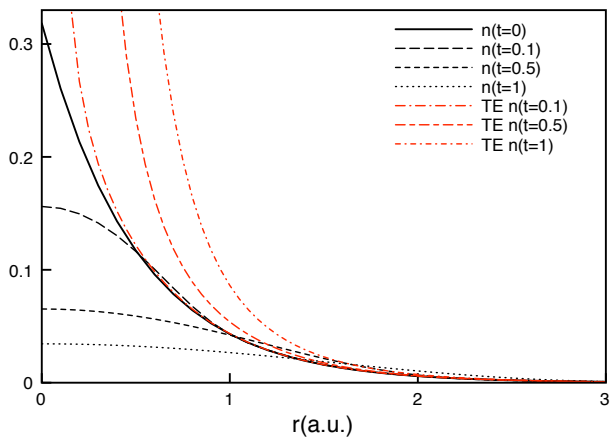}
\caption{TD exact density and TE density of the ground-state wavefunction of hydrogen atom under free-propagation.}
\label{fig:3dall}
\end{figure}

Begin with a simple example.  
Start from the ground-state wavefunction of the hydrogen atom, 
\ben
\label{eqn:3dHgs}
\psi_0(\br) = \frac{Z^{3/2}}{\sqrt{\pi}}\exp(-Zr).
\een
Our potential for $t\geq 0$ is simply zero:  The nucleus has been instantaneously
vaporized.
The exact TD wavefunction can be found by applying the
free-particle TD Green's function to $\Psi(0)$, yielding
\ben
\psi(\vect{r},t)=\frac{Z^{3/2}e^{iZ^2t/2}}{2\sqrt{\pi}r}[f(r,t)-f(-r,t)],
\een
where $f(r,t)=(r+iZt)\exp(Zr)\erfc[(r+iZt)/\sqrt{2it}]$.
Fig. \ref{fig:3dall} shows the exact solution and the power-series
solution as in Eq. (\ref{psiTE}), which can be summed to all orders:
\ben
\label{eqn:3dnucdisapp:TE}
\psi\TE(\vect{r},t)=\frac{Z^{3/2}}{\sqrt{\pi}}e^{-Z r+iZ^2t/2}(1-i\frac{Zt}{r}),\quad r>0.
\een
The TE density diverges as $r\to 0$ and so is invalid.


Our method begins with introducing reduced variables:
\ben
s=\frac{Z\mathrm{e}^2}{4\pi\epsilon_0\hbar}\sqrt{\frac{m_e t}{\hbar}},
\quad \brp=\sqrt{\frac{m_e}{2\hbar t}}\, \br
\een
where the powers are motivated by the previous example, and
constants chosen for convenience. These reduced variables are appropriate if the system contains only one nucleus placed at the origin.
In reduced variables and atomic units ($e^2=\hbar=m=1)$, the TDSE becomes
\ben
\left\{ {\cal L} -i s \partial/\partial s
+2s^2v/Z^2 \right\}\psi = 0,
\een
where 
\ben
{\cal L} = -\bar\nabla^2/2 + i \bar \br \cdot \bar\nabla,
\een
$Z$ is the nuclear charge, and $v$ is the external potential.
In the disappearing-nucleus case,
$\psi$ is equal to its Taylor-expansion in powers of $s$ for fixed $\brp$,
and thus we try:
\ben
\psi(\brp,s) = \sum_{p=0}^\infty\, \psi_p (\brp)\, s^p,
\een
yielding:
\ben
\label{eqn:psineqn}
\left\{ {\cal L} -m i \right\} \psi_m (\bar \br)
=-\frac{2}{Z^2}\sum_{p=0}^{m} v_{p-2}(\bar\br) \psi_{m-p} (\bar \br),
\een
where $v(\br,t)=\sum_p v_p(\bar\br)\, s^p$.  Each half-power of $t$ produces a second-order differential equation for
a function of $\brp$.  
The initial wavefunction in reduced variables is $\psi_0( {\sqrt{2}} \brp s/Z)$.  For finite
values of $\br$, as $s\to 0$, $\bar r \to \infty$,
so its expansion as $s\to0$ determines the
large $\bar r$ behavior of the $\psi_m(\brp)$.

To find the leading non-analytic behavior in $t$ due to a cusp, we rearrange the TE solution of Eq. (\ref{psiTE}) in terms of the reduced variables $s$ and $\brp$, and test its validity power by power in $s$.
Each $\psi\TE_m$  must satisfy the differential
equation, the boundary condition derived from the initial wavefunction as $\bar r \to\infty$, and remain finite at the
nucleus ($m$ being the order of the $s$-expansion).  We do this recursively, until we
find $k$, the lowest value of $m$ for which $\psi\TE_k$ fails.   Define
\ben
\xi(\bar\br, s) = \psi(\bar\br,s) - \psi\TE(\bar\br,s)
\een
and solve the differential equation for $\xi_k(\br)$ to find the leading non-analytic
behavior due to the cusp.  

This can be quite demanding as, in 3D, the equation is a partial differential
equation.  A simple approach is to perform a
local analysis using the method of dominant
balance\cite{BO99} to generate an asymptotic series in inverse powers of $\bar\br$.
Repeated application yields the series to all orders, but this is insufficient to
ensure that $\psi$ remains finite at the $r=0$.
This difficulty can be overcome by performing
a Borel resummation of the asymptotic series\cite{BO99} to find the exact solution.  
The conditions where Borel resummation works are discussed in \cite{S95}.

For the hydrogen atom in a suddenly switched electric
field which stays on indefinitely, the initial
wavefunction is that of Eq. \ref{eqn:3dHgs},
and the potential after $t=0$ is
\ben
v(r) = -Z/r + {\calE}\, z,
\een
where $\calE$ is the amplitude of the electric field.
This situation is covered by the Runge-Gross theorem as a potential that is analytic in
time (in fact, constant) with an initial wavefunction that is not its instantaneous ground state.
The time-dependent dipole moment $\mu(t)$ determines
the photoabsorption of the atom via
\ben
\sigma(\omega) = \frac{4\pi\omega}{c}\int_0^\infty\intd \tau\;\left[\left.\frac{\partial\mu(t)}{\partial t}\right|_{t=\tau}\right]\sin(\omega\tau),
\een
known analytically\cite{BS57,YFB09}
for the H atom as $\calE\to 0$.

To apply our method, we first calculate the TE wavefunction
order-by-order.  We find each term satisfies both the differential equation and boundary
conditions until the 4th order:
\begin{multline}
\psi_4\TE=\frac{Z^{3/2}}{24\sqrt{\pi}}(-3+12i{\bar r}^2+4{\bar r}^4)\\
\quad\quad+{\calE}\frac{i{\bar z}}{12\sqrt{\pi}Z^{3/2}{\bar r}^3}(1+6i{\bar r}^2+24{\bar r}^4)
\end{multline}
which works except that it diverges at the nucleus.   
Assume a correction of the form:
\ben
\xi_4 = \exp(S(\bar r))\, \bar z,
\een
because the applied field is linear in $z$.
This yields an ordinary differential equation for $S$:
\ben
S''+(S')^2-2i{\bar r}S'+{4S'}/{{\bar r}}+6i=0,
\een
and applying the method of dominant balance recursively, one finds
the asymptotic expansion:
\begin{multline}
\xi_4=c_1\left({\bar r}^3+\frac{9i{\bar r}}{2}-\frac{9}{4{\bar r}}+\frac{3i}{8{\bar r}^3}\right)\\
+c_2\frac{e^{i{\bar r}^2}}{{\bar r}^8}\left[1+\frac{1}{{\bar r}^2}\sum_{m=0}^\infty\frac{(-i)^{m+1}(m+4)(2m+6)!}{(m+1)!2^{2m+5}\cdot 9{\bar r}^{2m}}\right]
\end{multline}
Since $\psi\TE_4$ satisfies the boundary condition at large $\bar r$, $\xi_4$
must vanish here, and $c_1=0$.  But this asymptotic
series does not yield the small $\bar r$ limit.  To find this, we perform a
Borel resummation, yielding
\begin{multline}
\xi_4=
c_2\frac{1+i}{72{\bar r}^3}\Bigg[(2+2i)e^{i{\bar r}^2}{\bar r}(-3+16i{\bar r}^2+4{\bar r}^4)\\
-\sqrt{2\pi}(3i-18{\bar r}^2+36i{\bar r}^4+8{\bar r}^6)\erfc\left(\frac{1-i}{\sqrt{2}}{\bar r}\right)\Bigg]
\end{multline}
Requiring $\psi_4$ to remain finite yields
\ben
c_2 = (1-i){\calE}/(2\pi Z^{3/2}).
\een
One can easily check analytically that $\psi_4=\psi\TE_4+\xi_4$ is then the unique solution
to Eq. (\ref{eqn:psineqn}) for $k=4$.

With $\psi_4$, we find the
leading half-power of the wavefunction
by changing the variables back to $(\vect{r},t)$ and taking the asymptotic expansion
around the $t=0$. The leading time-half-power is then
\ben
\label{eqn:3dHinEfield:halfpower}
-\frac{(8-8i) {\calE} Z^{5/2}e^{ir^2/2t}\cos\theta}{\pi r^7}\,t^{11/2},\quad r\gg\sqrt{t},
\een
and one can show $\psi_{m>4}$ contributes to higher order time-half-powers by 
the method of dominant balance. This is also the lowest order in $t$ at which rapid non-time-analytic oscillations of the phase appear.

Our derivation applies for any value of $\calE$.  The one-body potential
appears linearly in our equations for $\psi_n(\brp)$, and only for $n \geq 3$.
Thus if $\psi_4(\brp)$ produces the leading non-analytic behavior in $t$, then
this behavior is linear in $\calE$, and is determined exactly by
linear response theory.  (Higher-order effects in $\calE$ produce changes in 
other terms, but not the leading non-analytic short-time behavior.) To see this, take the Lehmann representation of the Green's function
of the hydrogen atom
and evaluate the change in the wavefunction to first order in $\calE$. 
We did not find a closed-form, but the stationary phase approximation to the integral over the wave number yields
the leading half-power term 
in agreement with our short-time result Eq. (\ref{eqn:3dHinEfield:halfpower}).

The leading half-power term  in the induced dipole moment calculated using linear response theory yields:
\ben
\mu(t\to0_+) \sim \cdots-{256 Z^5}/({2835\sqrt{\pi}})\, t^{9/2}+\cdots.
\een
Its Fourier transform yields the known high-frequency decay of the photoabsorption cross-section\cite{BS57}
\ben
\label{eqn:photoabs}
\sigma(\omega\to\infty)\sim{16\sqrt{2}Z^5\pi}/({3c\omega^{7/2}}).
\een
Thus the cusp at the nucleus leads to fractional powers in time-evolution and
fractional powers in frequency decay.


The leading half-power term in $\sigma(\omega\to\infty)$ calculated with $\psi_4$ differs from Eq.~(\ref{eqn:photoabs}) by a factor of 2. In calculating $\sigma(\omega)$ from $\psi(\br,t)$, one must integrate over $r$, and the coupling between $r$ and $t$
in $\psi$ (as seen in $\psi_4$) 
allows higher order terms in the $s$-expansion to contribute to the leading half-power term in $\sigma(\omega\to\infty)$ as well. In this case, $\psi_4$ and $\psi_5$ both contribute to the $\omega^{-7/2}$ term in $\sigma$.

Before discussing the consequences for TDDFT, we again consider the simpler, but starker
example of the disappearing nucleus case. Our method correctly shows the error appearing
at $s^1$ order, and produces the correct short-time behavior:
\begin{multline}
\psi(\abs{\vect{r}}\gg\sqrt{t},t\to0_+)\sim\frac{Z^{3/2}e^{-Zr}}{\sqrt{\pi}}+\frac{iZ^{5/2}(Zr-2)e^{-Zr}}{2\sqrt{\pi}r}t\\
-\frac{Z^{9/2}(Zr-4)e^{-Zr}}{8\sqrt{\pi}r}t^2-\frac{(2+2i)Z^{5/2}e^{ir^2/(2t)}}{\pi r^4}t^{5/2}+\cdots
\end{multline}
while its TE contains only integer powers of $t$.

Although less relevant for TDDFT, our method applies equally well in 1d and
proves that the short-time behavior for the initial state with cusp example in \cite{MTWB10} originates from $s^1$ order
and has fractional power $t^{3/2}$, again with a rapid non-time-analytic phase.  Similarly, it's $s^4$ order for a particle in a delta-well with
an applied electric field, and the leading short-time behavior in $\psi$ is 
\ben
-\frac{(4+4i)\epsilon Z^{3/2}\exp[ix^2/(2t)]}{\sqrt{\pi}x^5}t^{9/2},\quad\abs{x}\gg\sqrt{t},
\een
consistent with the known polarizability and its high-frequency limit.\cite{YBxx}

Our results are proven only for single electrons.  However, the nuclear
potential dominates over the electron-electron repulsion in the region of the nucleus,
so that the cusp condition on the time-dependent density near a nucleus remains
valid regardless of the interelectron repulsion.  Thus we expect the qualitative features
(i.e., half-powers of $t$ in the short-time behavior) to remain true even in the
presence of interaction.  

We conclude by discussing the relevance of our 
results for TDDFT, whose theorems certainly apply (or not) for $N=1$. The proof of a
one-to-one correspondence\cite{RG84} shows
that, for two $t$-analytic potentials whose Taylor
expansions first differ in the $k$-th order,
the difference in the $k+2$th-order
$t$-derivative of the densities is non-zero. Thus
the two densities differ, the Runge-Gross proof applies, and the potential
is a functional of the density under the conditions stated\cite{RG84}.
Whether or not the TE density matches the true density is irrelevant.

On the other hand, the KS potential 
of Ref. \cite{L99} produces a density whose $t$-derivatives equal those
of the interacting density.
If we assume that the density is $v$-representable by a $t$-analytic potential 
(e.g., if the KS and interacting wavefunctions themselves are
$t$-analytic), then that potential will yield this density. 
But the wavefunctions are not $t$-analytic in the examples given here, and
typically are not when an initial wavefunction with a cusp undergoes a
non-trivial evolution. In general, a more sophisticated procedure is needed to
ensure the constructed potential generates the desired density,
and some progress has been made~\cite{MTWB10,T10,RL10}.
This is important as the constructive procedure has been invoked or
applied {\em as is} to a variety of
situations\cite{SPC10,YTRA10,RPB09,K09,LU08}.

One might argue that real atoms have finite nuclei, or that
real molecules and solids have nuclear wavefunctions that smear the
cusps due to nuclei.  But such arguments miss the basic point.
Just as in ground-state DFT,
time-dependent DFT is an exact mapping of the quantum mechanics of
electrons, for which there are no difficulties within BO
or need for finite nuclei.  

We thank Tchavdar Todorov for very useful discussions.
ZY and KB acknowledge U.S. D.O.E funding (DE-FG02-08ER46496) and
NTM acknowledges both NSF and the Research
Corporation. 


\end{document}